\documentclass[aps,prl,twocolumn,superscriptaddress]{revtex4}
\usepackage{graphicx,epsf}
\usepackage{amsmath}

\newcommand{\YRS}{YbRh$_2$Si$_2$}

\begin{document}

\title{
Reply to Comment by S. Friedemann {\em et al.} on\\
``Zeeman-Driven Lifshitz Transition: A Model for the
Experimentally Observed Fermi-Surface Reconstruction in \YRS''
}

\author{Andreas Hackl}
\affiliation{Department of Physics, California Institute of Technology, Pasadena, CA 91125, USA}
\author{Matthias Vojta}
\affiliation{Institut f\"ur Theoretische Physik, Technische Universit\"at Dresden,
01062 Dresden, Germany}

\date{\today}

\begin{abstract}
A reply to the comment by S. Friedemann {\em et al.} [arXiv:1207.0536] on our article
[Phys. Rev. Lett. {\bf 106}, 137002 (2011), arXiv:1012.0303].
\end{abstract}

\maketitle


%
In the light of experimentally observed anomalies in the temperature--field phase diagram
of the heavy-fermion metal \YRS, our Letter \cite{L} proposed a Zeeman-driven narrow-band
Lifshitz transition as an explanation which is alternative to the popular Kondo-breakdown
scenario. This idea was qualitatively illustrated utilizing simple toy-model calculations
for non-interacting electrons.
The Comment by Friedemann {\em et al.} \cite{C} argues that our ideas cannot apply to
\YRS\ because of them contradicting experimental data. Below we provide a point-by-point
reply in order to show that such contradictions do not exist.

First, we re-iterate \cite{L} that various measurements on \YRS\ point to the
existence of one or more small energy scales below the Kondo temperature, present even at
the transition field $B_c\approx 60$\,mT: distinct crossovers at different temperatures
below 0.5\,K are seen in the thermal expansion, in $\Gamma_{p,B}(T)$
\cite{kuechler03,tokiwa09}, the thermopower \cite{hartmann10}, and the thermal
conductivity \cite{wf11}.
The existence of these crossover scales limits straightforward interpretations in terms
of quantum criticality, and it partly motivated the proposal in Ref.~\onlinecite{L}.

Second, we note that, within the Lifshitz scenario, the small energy scale $E_c \sim
5\,\mu$eV (or 50\,mK) is actually {\em not} the total width of a narrow band, but the
distance between the Fermi level and the bottom (or top) of the relevant band piece
undergoing the Lifshitz transition, i.e., the depth of a Fermi pocket. The total width of
the band piece may be much larger, and the resulting peak in the quasiparticle density of
states (DOS) strongly asymmetric w.r.t. the Fermi level, see Fig.~1.

{\em A. Observability at elevated temperatures:}
A central point of the scenario of Ref.~\onlinecite{L} is that the properties for $T\gg
E_c$ are not that of the quantum critical regime of a Lifshitz transition, see Fig.~1 of
Ref.~\onlinecite{L}. Instead, in this regime the Fermi pocket is ``smeared'', and the
Zeeman splitting of the narrow band piece leads to apparent non-Fermi liquid behavior and
Schottky-like anomalies in thermodynamics.
Nevertheless, in the Hall effect, a distinct crossover occurring at $T_{\rm hall} \propto
B$ is still visible at temperatures $T\gg E_c$, as shown by explicit calculation in Fig. 3 of
Ref.~\onlinecite{L}.

{\em B1. Entropy crisis:}
This argument in Ref.~\onlinecite{C} does not apply, as the energy scale $E_c$ does
not replace the Kondo scale, but is an additional scale within the heavy-fermion band
structure, as is clearly stated on pg.~2 of Ref.~\onlinecite{L}.

{\em B2/B3. Weight of the DOS peak:}
The arguments in Ref.~\onlinecite{C} aim at constraining the weight of the narrow band
piece by a height (set by the specific heat coefficient $\gamma$) and a width (set by $E_c$).
While such considerations are valid in principle, care is required:
(i) Height: The narrow-band feature may well contribute a significant part of the $T\to0$
$\gamma$ value, which moreover is not known to good accuracy, because in our scenario it
will only be attained below $\approx 5$\,mK. E.g. $\gamma$ values of 5\,J/K$^2$mol appear
possible.
(ii) Width: As noted above, the DOS peak can be strongly asymmetric w.r.t. the Fermi
level, such that its width is much larger than $E_c$ (Fig. 1), and only a fraction of the
corresponding entropy is released at $T=50$\,mK.
Taken together, the weight estimate given in Ref.~\onlinecite{C} can be off by
an order of magnitude, and a weight corresponding to 1--2\% of $R\ln 2$
would be compatible with thermodynamic data.

The calculations in Ref.~\onlinecite{L} were not intended to quantitatively
match experiments (and hence did not take into account peak asymmetries etc.), and
we believe that a detailed quantitative comparison is not appropriate at this stage,
because correlation effects (e.g. the tendency towards ferromagnetism) are likely to
significantly influence the results.

{\em C. Hall crossover:}
Refs.~\cite{C,friede10} interpret the evolution of the Hall coefficient in terms of a
zero-temperature jump. This interpretation invokes an extrapolation to $T=0$, which is
problematic if small energy scales are present, see above. Also, sufficiently low
temperatures have not been probed: At 20\,mK, the crossover width is roughly 20\,mT which
is not small compared to $B_c$, i.e., even at the lowest investigated $T$ the crossover
is broad.
In fact, Fig.~3d of Ref.~\onlinecite{L} provided a proof of principle that a Lifshitz
scenario can be consistent with  the apparent linear-in-$T$ crossover width down to
20\,mK (without jump at $T=0$), i.e., there is no inconsistency.

{\em D. Enhancement of specific heat:}
Experimentally, the specific heat coefficient $\gamma$ is large near $B_c$ at the lowest
$T$; the claimed divergence is again based on a problematic $T\to 0$ extrapolation.
In our scenario, the large specific heat does {\em not} arise from the Lifshitz
transition per se, but from the presence of the narrow piece of heavy-fermion band which
causes the peak in the quasiparticle DOS: $\gamma$ keeps increasing upon
lowering $T$ until $T$ becomes smaller than the energy scale on which the DOS varies --
there is no inconsistency with experimental data for $B\geq B_c$.

In the narrow range $B<B_c$ the detailed shape of the DOS peak becomes relevant, and an
asymmetric peak could explain a decreasing $\gamma$ upon lowering $B$. More importantly,
the interplay with antiferromagnetism needs to be considered which is beyond the scope of
Ref.~\onlinecite{L}.

We note that the weak singularities of a Lifshitz transition, alluded to in
Ref.~\onlinecite{C}, will only be relevant at ultra-low temperatures (below 10\,mK), see
Fig.~1 of Ref.~\onlinecite{L}.

{\em E. Role of antiferromagnetism.}
It is true that antiferromagnetism (AF) is not contained in the Lifshitz scenario, and is
assumed to be secondary. We believe that this assumption is not in contradiction to
experiments. On the contrary: It has been shown that AF and the crossovers associated
with the so-called $T^\ast$ line \cite{gegenwart07} can be separated by doping
\cite{friede09}. This is most striking in recent data on \YRS\ with 5\% Fe doping, where
the $T^\ast$ line signatures are seen disconnected from any AF transition \cite{geg}.
A plausible idea would be that AF and the Lifshitz transition are two separate
phenomena emerging from different portions of the Fermi surface.

In summary, we feel that the scenario of Ref.~\onlinecite{L} continues to be viable
candidate to explain salient features of the $T-B$ phase diagram of \YRS. A
more quantitative modeling needs to treat correlation effects beyond assuming
effective quasiparticle band structures; initial work in this direction is in progress.


\begin{figure}
\includegraphics[width=0.48\textwidth]{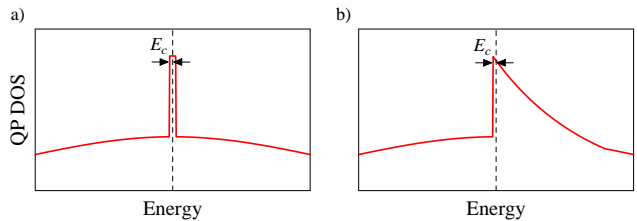}
\caption{
Quasiparticle DOS (schematic), with a symmetric (a) or strongly asymmetric (b) narrow
peak on top of the heavy-fermion background. $E_c$ is the distance between the Fermi
level (dashed) and the energy where the DOS strongly decreases (i.e. the bottom of the
assumed weakly dispersing piece of band); in case (b) the effective width of the peak is
much larger than $E_c$.
}
\end{figure}



\begin{thebibliography}{}

\bibitem{L}
A. Hackl and M. Vojta,
Phys. Rev. Lett. {\bf 106}, 137002 (2011).

\bibitem{C}
S. Friedemann {\em et al.},
preceeding comment, preprint arXiv:1207.0536.

\bibitem{kuechler03}
R. K\"uchler {\em et al.},
Phys. Rev. Lett. {\bf 91}, 066405 (2003).

\bibitem{tokiwa09}
Y. Tokiwa {\em et al.},
Phys. Rev. Lett. {\bf 102}, 066401 (2009).

\bibitem{hartmann10}
S. Hartmann {\em et al.},
\prl {\bf 104}, 096401 (2010).

\bibitem{wf11}
H. Pfau {\em et al.}, Nature {\bf 484}, 493 (2012).


\bibitem{friede10}
S. Friedemann {\em et al.},
PNAS {\bf 107}, 14547 (2010).

\bibitem{gegenwart07}
P. Gegenwart {\em et al.},
Science {\bf 315}, 969 (2007).

\bibitem{friede09}
S. Friedemann {\em et al.},
Nature Phys. {\bf 5}, 465 (2009).

\bibitem{geg}
Y. Tokiwa and P. Gegenwart, unpublished results.

\end{thebibliography}
\end{document}